\documentstyle[12pt,epsfig]{article}  
\begin{document} 
\baselineskip=20pt

\def\la{\mathrel{\mathpalette\fun <}}
\def\ga{\mathrel{\mathpalette\fun >}}
\def\fun#1#2{\lower3.6pt\vbox{\baselineskip0pt\lineskip.9pt
\ialign{$\mathsurround=0pt#1\hfil##\hfil$\crcr#2\crcr\sim\crcr}}} 

\begin{titlepage} 
\begin{center}
{\Large \bf  Jet activity versus alignment }

\vspace{4mm}
I.P.~Lokhtin$^a$, A.K.~Managadze$^b$, L.I.~Sarycheva$^c$, A.M.~Snigirev$^d$  \\
M.V.Lomonosov Moscow State University, 
D.V.Skobeltsyn Institute of Nuclear Physics, \\
119992,  Vorobievy Gory, Moscow, Russia \\

\end{center}  

\begin{abstract} 
The hypothesis about the relation between the observed alignment of spots in 
the x-ray film in cosmic ray emulsion experiments  and the features of events 
in which jets prevail at super high energies is tested. Due to strong 
dynamical correlation between  jet axis directions and that between 
momenta of jet particles (almost collinearity), the evaluated degree of 
alignment is considerably larger than that for randomly selected  chaoticly 
located spots in the x-ray film. It appears comparable with experimental data 
provided that the height of primary interaction, the collision energy and the 
total energy of selected clusters meet certain conditions. The Monte Carlo 
generator PYTHIA, which basically well
describes jet events in hadron-hadron interactions, was used for the analysis.

\end{abstract}

\bigskip


\vspace{100mm}
\noindent
------------------------------------------------------\\
$^a$e-mail: igor@lav01.sinp.msu.ru \\
$^b$e-mail: mng@dec1.sinp.msu.ru \\
$^c$e-mail: lis@alex.sinp.msu.ru \\ 
$^d$e-mail: snigirev@lav01.sinp.msu.ru \\
\end{titlepage}   
\newpage 

\section{Introduction}

The intriguing phenomenon, the strong collinearity of cores in emulsion
experiments~\cite{pamir}, closely related to  coplanar scattering of 
secondary particles in the interaction, has been observed  long time ago. So 
far there is no simple satisfactory explanation of these cosmic ray 
observations in spite of numerous attempts
to find it (see, for instance,~\cite{book} 
and references therein). Among them, the jet-like mechanism~\cite{halzen} looks 
very attractive and gives the natural explanation of alignment of three spots 
along the straight line which results from momentum 
conservation in a simple parton picture of
scattering. Besides, the strong momentum correlation of particles inside a jet
and correlation between 
jet axes due to  singularity of QCD matrix elements allow us to
suggest the  high degree of alignment for more than 3 spots. This has
been already demonstrated for the four cores in~\cite{halzen} but using a
simplified picture of hadronization.

With increasing energy  of colliding hadrons (nuclei)
hard and semi-hard jets begin to play an important role due to  growth of 
their production cross sections. Thus the jet activity is likely to be a feature
of all events above certain threshold collision energy. One of the
manifestations of this activity and the strict momentum 
ordering inside a hard enough jet can be the observed strong collinearity of 
spots in emulsion experiments. The main purpose of the  
present paper is just to trace this relation in detail. 
In Sect. 2 we formulate the problem  on the whole. Section 3  describes 
the results of numerical simulation made under conditions 
close to emulsion experiments in the framework of PYTHIA~\cite{pythia}, and  
some discussion. A summary can be found in Sect. 4. 

\section{Problem under consideration}

In the Pamir experiment the observed events ($\gamma$-hadron families with the
alignment) are produced, mostly, by a  proton with  energy  $\ga 10^4$ TeV
interacting at a height of several hundred metres to several kilometres
in the atmosphere above the chamber~\cite{book,rak}. 
The collision products are observed within a 
radial distance up to several centimetres in the emulsion where the spot 
separation is of the order of 1 mm. One can estimate the typical transverse 
momentum in the events under consideration using the ratio 
(see also (\ref{position})):
\begin{equation} 
\label{ratio}
p_T h ~=~r E, 
\end{equation} 
where $E$ is the energy deposition in the spot, 
$r$ is its spacing in the $x$-ray film,
$h$ is the height of interaction. So $p_T$ is of 
the  order of 10 GeV for $r=15$ mm, 
$h=1$ km, $E=700$ TeV. The particles with such transverse momenta can be
typically initiated by a jet with $p_T^{\rm jet} \sim 50$ GeV or larger, 
since the most probable value of a fraction of jet energy carried by leading 
particles is $\sim 0.2$~\cite{pythia}. Such energetic jets are already enough 
collimated: their effective angular cone size $\theta_{\rm eff} \la 15^{\circ}$ 
due to the strict ordering of transverse and longitudinal particle momenta in 
the leading logarithm
approximation of perturbative QCD~\cite{dok}. QCD  ``teaches" us also that this 
effective size decreases with the growth of the ``jet hardness" (transverse 
momentum) as
\begin{equation} 
\label{hardness}
\theta_{\rm eff}~ \sim ~ (\ln(p_T^{\rm jet}/\Lambda_{\rm QCD}))^{-1} , 
\end{equation} 
where $\Lambda_{\rm QCD}$ is the dimensional QCD parameter.

The main conjecture is that the  particle distribution from 
the hard enough jets  
in the $x$-ray film plane can lead to the alignment of the emulsion spots due to
the strong collimation of such particles and dynamical correlation between  jet 
axis directions. Let us consider the kinematics
in  detail. For our analysis it is convenient to parametrize 4-momentum of each
produced particle $i$ under consideration with its transverse momentum $p_{Ti}$
(relative to the collision axis $z$), azimuthal angle $\phi_i$ and rapidity
$\eta_i$ in the center-of-mass system:  
\begin{equation} 
\label{momentum}
 [\sqrt{p^2_{Ti}+m^2_i}~\cosh \eta_i,~~~ p_{Ti}\cos \phi_i,~~~
p_{Ti}\sin \phi_i,~~~ \sqrt{p^2_{Ti}+m^2_i}~\sinh \eta_i]. 
\end{equation} 
In this case the transformation from the center-of-mass 
system to the laboratory one
reduces to a simple rapidity shift: $\zeta_i=\eta_0+\eta_i$, where
$\zeta_i$, $\eta_0$ are the rapidities of particle $i$ and  the center-of-mass 
system correspondingly in the laboratory reference frame. If we neglect  the further
interactions of particles  propagating through the atmosphere (this
gives the maximum estimation of the alignment effect), then their 
position in the transverse $(xy)$-plane is easily calculated  
\begin{equation} 
\label{position}
{\bf r}_i~=~ \frac{{\bf v}_{ri}}{v_{zi}}~h~=\frac{{\bf p}_{Ti}}
 {\sqrt{p^2_{Ti}+m^2_i}~\sinh (\eta_0+\eta_i)}~h~,
\end{equation} 
where ${\bf v}_{ri}$ and $v_{zi}$ are the radial and longitudinal components of 
particle velocity respectively 
($E_i~=~ \sqrt{p^2_{Ti}+m^2_i}~\cosh (\eta_0+\eta_i)$ is the particle energy in
the laboratory frame). 

Since the size of the observation region is of the order of several centimetres,
these radial distances must  obey the following restriction:
\begin{equation} 
\label{mini}
 r_{\rm min}~<~r_i,
\end{equation} 
\begin{equation} 
\label{max}
 r_{i}~<~r_{\rm max}.
\end{equation} 
We set $r_{\rm min}=r_{\rm res}\simeq 1$ mm, ~~ $r_{\rm max}\simeq 15$ mm.  The
restriction (\ref{mini})  simply means that  spots are not mixed with the
central one formed by the particles which fly close to the collision axis (mainly
from the fragmentation region of an incident proton). The separation of spots
in the $x$-ray film gives another restriction on the  distance between
particles
\begin{equation} 
\label{dij}
d_{ij}~=~ 
 \sqrt{r^2_i~+~r^2_j~-~2r_i r_j \cos(\phi_i~-~\phi_j)}. 
\end{equation} 
It must be larger than 1 mm:
\begin{equation} 
\label{dijres}
d_{ij}~>~ r_{\rm res}~, 
\end{equation} 
in the opposite case the particles must be combined in one particle-cluster
until there remain only particles and/or particle-clusters   with the mutual 
distances larger than $r_{\rm res}$, each such 
particle-cluster being considered as
a single particle with coordinates defined in the same way as
center-of-mass coordinates of two bodies:
\begin{equation}
\label{rij}
{\bf r}_{ij}=({\bf r}_i E_i+ {\bf r}_j E_j)/(E_i+E_j). 
\end{equation}
Then we select $2,...,7$ clusters/particles which are most energetic and
obey the restrictions (\ref{mini}, \ref{max}, \ref{dijres}) and calculate the
alignment $\lambda_{N_c}$ using the conventional definition~\cite{book,rak}: 
\begin{equation} 
\label{alig}
\lambda_{N_c}~=~ \frac{ \sum^{N_c}_{i \neq j \neq k}\cos(2 \phi_{ijk})}
{N_c(N_c-1)(N_c-2)}, 
\end{equation} 
and taking into account the central cluster, i.e. $N_c-1=2,...,7$.
Here $\phi_{ijk}$ is the angle between two vectors ($\bf {r}_k-\bf{r}_j$) 
and ($\bf{r}_k-\bf{r}_i$) (for the central spot $\bf{r}=0$).
This parameter characterizes  the location of $N_c$ points just along the
straight line and varies from $-1/(N_c-1)$ to $1$. For instance, in the case
of the symmetrical and close to  most probable random configuration of three
points in  a plane (the equilateral triangle) $\lambda_3 = -0.5$.                  
The ultimate case of perfect alignment is $\lambda_{N_c}=1$ 
when all points lie exactly along the 
straight line, while for an isotropic distribution $\lambda_{N_c} < 0$. The
alignment degree $P_{N_c}$ is defined as a fraction of events with
 $\lambda_{N_c} > 0.8$~\cite{book} with the
 number of cores not less than $N_c$.

\section{Numerical results and discussion}

If the hypothesis about the relation of alignment to the prevailing 
jet character of events at super high energies is valid, 
then this must manifest itself  first of all in nucleon-nucleon collisions. 
Therefore, to be specific we consider a collision 
of two protons  and fix a primary energy in the laboratory
system $E_{\rm lab} \simeq 9.8 \times 10^4$ TeV, that is equivalent to 
$\sqrt{s}\simeq 14$ TeV --- just the energy attainable at LHC  (the rapidity
shift being $\eta_0\simeq 9.55$ after the transformation from the 
center-of-mass system to the laboratory one). To simulate a collision of two
protons with  such energies we use the Monte Carlo generator
PYTHIA~\cite{pythia}, which basically well  describes jet events in 
hadron-hadron interactions and is tuned using the available experimental
accelerator data.

The results of numerical simulation which follows the consideration
in the previous Section are presented in Fig. 1 (solid curve) with the
parameters $r_{\rm min}=r_{\rm res}=1$ mm, $r_{\rm max}=15$ mm, $h=1000$ m,
which are close to the conditions of emulsion experiments, with the additional
restriction on the energy threshold of particle registration in the emulsion:
$E_i > E^{\rm thr}=4$ TeV \footnote{Our calculations are practically
insensitive to this threshold in the wide interval of its varying.}.
The estimated alignment degree $P_{N_c}$ for $N_c$ cores is considerably larger
than that for randomly selected  chaoticly located spots in the $x$-ray 
film, but is still too small (by a factor of 3---4) to describe the experimental 
data~\cite{man} even taking into account their large errors. This can mean
that the jet activity is not sufficient 
at such energies or the jet mechanism can
not, in principle, give the large experimentally observable alignment.

In order to try to answer  this question let us consider the influence of  
the applied restrictions (\ref{mini}, \ref{max}) (the laboratory acceptance 
criterion) on the spectrum of particles selected to calculate the alignment. For
particles with high enough transverse momenta $p_{Ti}$ relative to their
masses $m_i$  these conditions  (\ref{mini}, 
\ref{max}) reduce, mainly, to the restriction on the available particle
rapidities in the center-of-mass system:  
\begin{equation} 
\label{mini1}
 r_{\rm min}<r_i \Longrightarrow \eta_i < \eta_{\rm max} =
 \ln(r_0/r_{\rm min})\simeq 4.95,
\end{equation} 
\begin{equation} 
\label{max1}
 r_{i} < r_{\rm max} \Longrightarrow \eta_i > \eta_{\rm min} =
 \ln(r_0/r_{\rm max})\simeq 2.25,
\end{equation}
since in this case $r_i \simeq r_0/e^{\eta_i}$ for
$\eta_0 + \eta_i \ga 1$, where
\begin{equation}
\label{r0} 
r_0~=~2h/e^{\eta_o}.
\end{equation}  
Due to the kinematical restriction~\cite{dok},
\begin{equation} 
\label{kin}
e^{\eta^{\rm jet}}p_T^{\rm jet}\la \sqrt{s},
\end{equation}         
a production of harder jets with larger rapidities
becomes possible with the growth of $\sqrt{s}$. 
The rapidity region (\ref{mini1}, \ref{max1}) just corresponds to the
transition  from soft to hard QCD physics, where the jet activity {\it
could} manifest itself.

Here one should note that ultrarelativistic particles ($p_{Ti} \gg m_i$) 
are detected in the $x$-ray film from the restricted rapidity region 
(\ref{mini1}, \ref{max1}) which excludes such
configurations as back-to-back hard jets 
with rapidities close to zero in the center-of-mass system. 
But just such configurations with scattering of hard partons at angles close to
$90^{\circ}$ in the considered hadronic center-of-mass system (which in this
case practically coincides
with the  partonic center-of-mass system)
can be expected to be responsible for the alignment phenomenon.
The point is that leading particles from   both these hard jets have 
quantitatively comparable energies in the laboratory frame together with the 
``strong memory'' of scattering plane. Meanwhile leading particles from any 
other back-to-back hard jets with the relatively large modulo
rapidities $|\pm \eta^{jet}_{p.c.m.s.}|$ in the partonic
center-of-mass system have essentially different laboratory energies 
due to  Lorentz 
boost. And the energy distinction is mainly determined by the value of
$\exp(2|\pm \eta^{jet}_{p.c.m.s.}|)$. In the latter case
particles from a forward hard jet produce as a rule the most energetic clusters
(apart from the central one) in the 
laboratory frame, even if the 
particles from a backward (in the partonic center-of-mass system) jet
hit the detection region. However such most energetic clusters from one jet
are less correlated with the primary scattering plane and therefore will not 
be much aligned as clusters from both hard jets.
This argumentation is confirmed by our simulation. 

In this connection it is necessary to comment on the work~\cite{halzen} in which
the jet hypothesis has been suggested for the first time for the explanation 
of alignment phenomenon. There  the high degree of alignment has been
demonstrated 
for the four cores only, using a simplified picture of fragmentation 
process. In fact, an axis distribution has been calculated in the partonic 
center-of-mass system in the first order of perturbative QCD
theory at the partonic level, considering three partons in the final state only. 
Then the Lorentz transformation has been done in order to find their 
directions and localizations with respect to the central spot 
in the laboratory frame, considering each parton-jet as one long-living 
system (the fragmentation  time is of the order of flight time)
with some  effective mass and  aggregate group velocity. The velocity has been
fixed ($\beta^{\star}\simeq0.7$ in~\cite{halzen}) so as to be able to include
the events giving the high degree of alignment and corresponding to the two
final-state parton-jets in the backward hemisphere and one in the forward
hemisphere in the partonic center-of-mass system. In our variables this 
means  that a mass
factor $p_{T}/{\sqrt{p^2_{T}+M^2}}$ must be very small for a such massive
system ($M\gg p_{T}$) so that it  hits  the detection region even
with negative rapidities  (i.e. corresponds to the backward hemisphere
in the partonic center-of-mass system), 
if one uses the same boost parameters as for the 
hadronic center-of-mass system without taking into account the possible
additional boost due to the distinction between these partonic and hadronic
frames. Note that for real particles, e.g. $\pi$-mesons which mainly
contribute to the multiplicity, this mass factor 
becomes significant for very small transverse momenta, $p_{T \pi} 
\ll m_{\pi}=0.14$ GeV, only. However, as our investigation shows, 
falling of appropriately correlated particles into the observation region 
is still not sufficient to obtain the  high degree of alignment because of the 
energy selection procedure, if the total number of particles is 
large  and they generate many distinctly separated spots.

For completeness one should also mention that high transverse momentum jet 
production has a connection to the double-core configuration 
of cosmic-ray events as it has been pointed out in~\cite{cline}. 
Under certain conditions a hard forward (in the partonic center-of-mass system)
jet together with a central bunch gives two relatively far separated clusters
with large energies. The detailed studies of
double-core (or binocular) phenomena with estimations of event rates and average
lateral spread of the $\gamma$-family using a PQCD based Monte Carlo 
can be found, for instance, in~\cite{cao1, cao2}.

Ultrarelativistic particles from the central rapidity region in the hadronic
center-of-mass system  (as possible sources of appropriately correlated
spots) can hit the observation region  owing to the
decrease of  $r_0$ {\it only}, i.e. the  decrease of the  height $h$ of primary
interaction  or the increase of the rapidity  $\eta_0$ of the center-of-mass 
system due to the growth of energy $\sqrt{s}$, as it follows from (\ref{r0}).
The energy growth seems preferable, if we intend to be  closer to  emulsion
experiments and  increase the jet activity. However this demands the
extrapolation of PYTHIA parameters and their special tuning to the 
experimentally untested energy domain. Updating can be done appropriately after
the LHC operation starts. 
Moreover at present  this generator already uses  the extrapolation 
of experimentally tested cross sections and structure functions 
to the LHC  energy
region $\sqrt{s}\simeq 14$ TeV in order to estimate the effects expected at such
energies.

For illustration we utilize the first ``less dangerous'' alternative  ---
decrease the interaction height by a factor of $20$ rather than increase
the energy $\sqrt{s}$ by the same factor of $20$ at the initial height
so that particles from both hard jets
(with back-to-back structure),  hitting  the 
registration region, come from some rapidity
range near $\eta_i \simeq 0$ including adjoint positive and negative 
values. In this case the alignment degree becomes strongly
dependent on the minimum transverse momentum of hard process, $p_T^{\rm hard}$, 
which is a parameter of PYTHIA. At the height $h=1$ km such dependence was not
visible, although we might catch some marginal tendency  of the alignment degree
to grow  with the increase of $p_T^{\rm hard}$ at that height. However
without the restriction on $p_T^{\rm hard}$ from below (minimum bias) the result
coincides practically with one obtained earlier (solid curve in Fig. 1) that 
shows some general characteristics of jet structure of events. If $p_T^{\rm jet} 
\geq  3$ TeV, particles from these hard jets 
together with particles flying close to $z$-axis (within the 
transverse radius $< 1$ mm) result in the alignment degree (dashed
curve) comparable with the experimentally observed one~\cite{man}.

Thus the jet-like mechanism can, in principle,  attempt to 
explain the results of  emulsion experiments. For such an explanation
it is necessary (but not sufficient) 
that particles from both hard jets (with rapidities 
near $\eta_i \simeq 0$ in the center-of-mass system)  hit   the observation
region. This is possible at the relatively small height $h=50$ m and
$\sqrt{s}\simeq 14$ TeV; or at the height $h=1000$ m, but the considerably higher
energy $\sqrt{s}\simeq 14 \times 20 = 280$ TeV; or at some reasonable and 
acceptable intermediate combination of $h$,  $\sqrt{s}$ 
and $r_{\rm max}$ which meets the following condition:
\begin{equation}
\label{r00} 
r_0 = 2h/e^{\eta_o} = 2h m_p/ \sqrt{s} \la k r_{\rm max},
\end{equation} 
where $m_p$ is the proton mass. 
$k \simeq 1/2 < 1$ is needed in order to have particles with $\eta_i < 0$ that
hit the detection region (see (\ref{max1})).
We verified the decisive significance
of condition (\ref{r00}) to allow the observation
of large degree of alignment and its dependence on the 
process hardness  for the smaller energy $\sqrt{s}\simeq 1.4$ TeV (where the
prediction of PYTHIA is quite adequate) and  the height 
$h=5$ m  (in accordance with (\ref{r00})) thereby
confirming this peculiar kinematic ``scaling''.

At $p_T^{\rm hard} = 3$ TeV jets carry away about half of the energy of 
colliding protons in the center-of-mass system due to the relationship 
in a parton picture $\xi \simeq 2p^{\rm jet}_T/ \sqrt{s}$, where $\xi$ 
is a fraction of proton energy carried by each interacting 
parton (quark or gluon). The
striking feature of such configurations in the $x$-ray film is  
approximate equality of energy deposition in the central and the rest
most energetic clusters, that can be one of the physical guideline to select 
the events with very hard jets not only at the generator level (simulation). If we
simply apply the additional threshold on the minimum energy of detected clusters
needed in the alignment analysis, then we still obtain neither the desirable
selection of jet hardness nor the increase of the alignment degree. The small
variation of resolution parameter $r_{\rm res}$ does not provide the desirable
effect also. However,  introduction of   another  threshold  on the total 
energy of all $(N_c-1)$ selected clusters 
$ E^{\rm thr}_{\Sigma}\sim E_{\rm lab}/2$ (without taking into account the 
energy deposition in the central cluster around $r=0$),
\begin{equation}
\label{Ethr} 
\sum^{N_c -1}_{l=1}E_l > E^{\rm thr}_{\Sigma},
\end{equation}  
allows us to select the events with hard jets only in a ``natural'' physical
way and to reduce the hypothesis to the really active mechanism. 
Figure 2 shows that the alignment degree increases 
with the growth of $E^{\rm thr}_{\Sigma}$
(the restriction on $p_T^{\rm hard}$ is absent at all!), and it
becomes large enough (dashed curve) and comparable with the experimentally
observed one~\cite{man} above the threshold 
$ E^{\rm thr}_{\Sigma}\simeq 0.1 E_{\rm lab}\simeq 10$ PeV. Though one should
note that our estimations give still too steep dependence on $N_c$ as one
can see in Figs. 1b, 2b from comparison of slopes of straight lines with the
experimental behaviour.

To give the reader a feeling for the
various measures of alignment we present in Figs. 3 and 4 the spatial 
distributions of most energetic clusters in 
the $(xy)$-plane for a few generated events along with the
corresponding values of $\lambda_{N_c}$. Some spots are hardly visible because
of  their small sizes which are proportional to the cluster energies
(especially in the case  $\lambda_4 > 0.8$)
or because they are outside a square $10\times10$ mm$\times$mm
(but inside a circle $r=15$ mm) as it  sometimes happens in the case 
$\lambda_8 > 0.8$.
Besides for $\lambda_4 > 0.8$ we can  distinctly see three relatively
large spots resulted from two hard jets and a central bunch.

Here one should note that  there was slightly other
criterion for the selection of families for the  analysis in the works of Pamir
Collaboration: the families with the total energy of $\gamma$-quanta larger 
than a certain threshold and at least one hadron present 
were selected and analyzed.
The alignment becomes apparent considerably at $\sum E_{\gamma} > 0.5$ PeV
(the families being produced, mostly, by a  proton with  energy 
$\ga 10$ PeV). Since the adequate comparison of  our estimations with 
experimental data is impossible without a full simulation of particle 
propagation  
through the atmosphere, taking into account the energy distribution of primary
cosmic particles, etc., then in order to demonstrate  the
possibility of appearance of high alignment degree 
due to the jet mechanism we 
restrict ourself to the simpler (but as concerns physics essentially  close to 
experiment)
criterion of selection over the total energy of all particles.  These
particles are mostly $\pi$-mesons, the neutrals among them being the main source of the
detected $\gamma$-quanta. It is natural that the threshold on the total energy
of all particles must be larger than the similar  threshold on the total 
energy of 
$\gamma$-quanta at the same collision energy. For comparison we estimate also
the alignment degree selecting only the most energetic
$\gamma$-quanta with their total energy 
larger than certain threshold $E^{\rm thr}_{\gamma}$ (Fig. 5): 
\begin{equation}
\label{Egamma} 
\sum^{N_c -1}_{l=1}E_{l\gamma} > E^{\rm thr}_{\gamma}.
\end{equation}  
The result is close to that obtained  previously with the threshold 
imposed on the total energy of all particles. 

Besides for jet events 
\begin{equation}
\label{Pn} 
\frac{P_{N_c}}{P_{N_c+1}}~=~{\rm const}
\end{equation}  
with a high accuracy (see Figs. 1b, 2b, 5b, which
present the dependence of alignment
degree on the number of considered cores at the different values of hardness
parameter (1b) and threshold total energy (2b, 5b)  
in the logarithmic scale).
This constant depends on $p_T^{\rm hard}$, $E^{\rm thr}_{\Sigma}$, 
$E^{\rm thr}_{\gamma}$, decreasing
with their growth, and could in principle be determined by the kernels of the
Gribov---Lipatov---Altarelli---Parisi---Dokshitzer equations~\cite{dok, grib, 
lip, dok2, altar}  which describe the process
of radiation of quarks and gluons in the initial and final states. And, in fact,
this process is implemented in the PYTHIA generator together with the
subsequent hadronization of quarks and gluons.

If nevertheless particles from the central rapidity region $\eta_i \simeq 0$
and the jet-like mechanism are insufficient
to describe the observed alignment and there is another
mechanism of its appearance at the energy  $\sqrt{s} \sim 14$ TeV and the
height $h \sim 1000$ m (mostly used in emulsion experiment estimations), then
in any case some sort  of alignment should arise at LHC
too in the rapidity region 
(\ref{mini1}, \ref{max1}). This region must be investigated more carefully
on the purpose to study the azimuthal anisotropy of energy flux in accordance
with the procedure applied in the emulsion and other experiments, i.e. one
should analyze the energy deposition in the cells of $\eta \times \phi$-space
in the rapidity interval (\ref{mini1}, \ref{max1}) (the equivalent threshold 
minimum particle energy being $E^{\rm thr}_{\rm c.m.s.}=E^{\rm thr}/
\cosh \eta_0\simeq 2E^{\rm thr}/ e^{\eta_0}\simeq 0.6$ GeV 
in the center-of-mass system). Note that the absolute 
rapidity interval can be shifted:
it is necessary only that the difference ($\eta_{\rm max}-\eta_{\rm min})$
is equal to $\simeq 2.7$ 
in accordance with the variation of radial distance by a factor of 15  
($r_{\rm max}/r_{\rm min}= 15$) due to the relationship 
$r_i\simeq r_0/e^{\eta_i}$ (independently of $r_0$). In other words, since we use
particle momenta in the center-of-mass system, then future data
should be treated in accordance with the algorithm described earlier in Sects. 2,
3 introducing the corresponding laboratory observables.

\section{Conclusions}

Our analysis shows that for $pp$-collision at a fixed height of primary
interaction above the  energy $\sqrt{s}$, when the condition 
(\ref{r00}) is fulfilled --- that is  ultrarelativistic
particles from 
the rapidity interval near $\eta_i \simeq 0$  in the center-of-mass system 
fall into  the observation region
inside the radius $r_{\rm max}$ in the laboratory frame due to the large
Lorentz factor ---   the alignment of spots arises (this, in
principle, explains the existence of the experimental 
energy threshold of this effect) 
and the alignment degree becomes  strongly dependent  on
the process hardness. If the process hardness is close to  maximum for
the given energy $\sqrt{s}$, the estimated degree of alignment is already
comparable with the experimentally observed one. Introducing another additional
threshold (the scale of which is determined by the energy of an incident proton) on
the total energy of all $(N_c-1)$ selected most energetic clusters (without
taking into account the energy deposition in the central cluster) allows us to
select the events with  high hardness in a ''natural'' physical way
and thereby support the jet-like hypothesis, which later on  may be 
accepted (or  refuted) in further investigations of, for instance, the energy
cluster distribution   and their particle composition with regard for
interactions in the atmosphere,  etc.

Meanwhile we suggest the more careful investigation of  the rapidity 
region (\ref{mini1}, \ref{max1}) at LHC  in order to reveal  the new 
still unknown  mechanisms of alignment if they exist. 
For this purpose  one should perform the analysis
of energy deposition in calorimeters of  CMS and ATLAS experiments
in accordance with the procedure described in Sects. 2, 3 (i.e. calculating the
appropriate observables in the laboratory frame). Such investigation can clarify
the origin of the alignment, test the alternative hypotheses and give the new
restrictions on the values of height and energy. 
 
\noindent
{\it Acknowledgements.} 

\noindent
It is pleasure to thank A.I. Demianov, S.V. Molodtsov, S.A. Slavatinsky, 
L.G. Sveshnikova, K.Yu. Teplov and G.T. Zatsepin for  discussions. 
This work is supported 
by grant N 04-02-16333 of Russian Foundation for Basic Research.

\begin{figure} 
\begin{center} 
\makebox{\epsfig{file=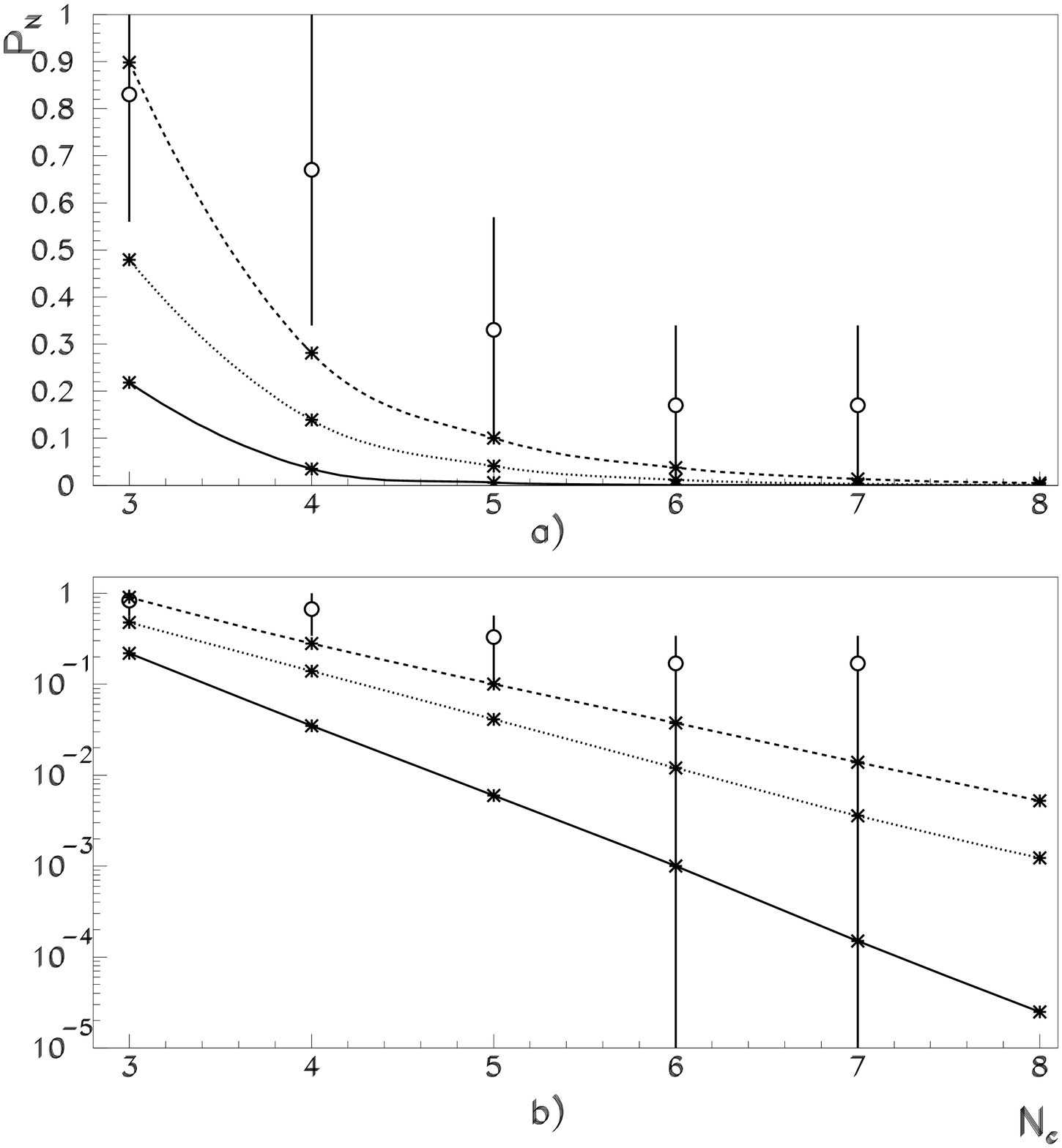, height=220mm}}    
\caption{\small The alignment degree $P_{N_c}$ as a function of cluster number $N_c$ 
at $h=50$ m and  $\sqrt{s}=14$ TeV in linear (a) and logarithmic (b) scales. The 
solid curve is the result (coincident with one at $h=1000$ m)
 without restriction on the minimum value of process
hardness $p^{\rm hard}_T$, the dotted curve --- at $p^{\rm hard}_T=300$ GeV,
the dashed curve  --- at $p^{\rm hard}_T=3$ TeV. Points
($\circ$) with errors are experimental data from~\cite{man}.}
\end{center}
\end{figure}

\begin{figure} 
\begin{center} 
\makebox{\epsfig{file=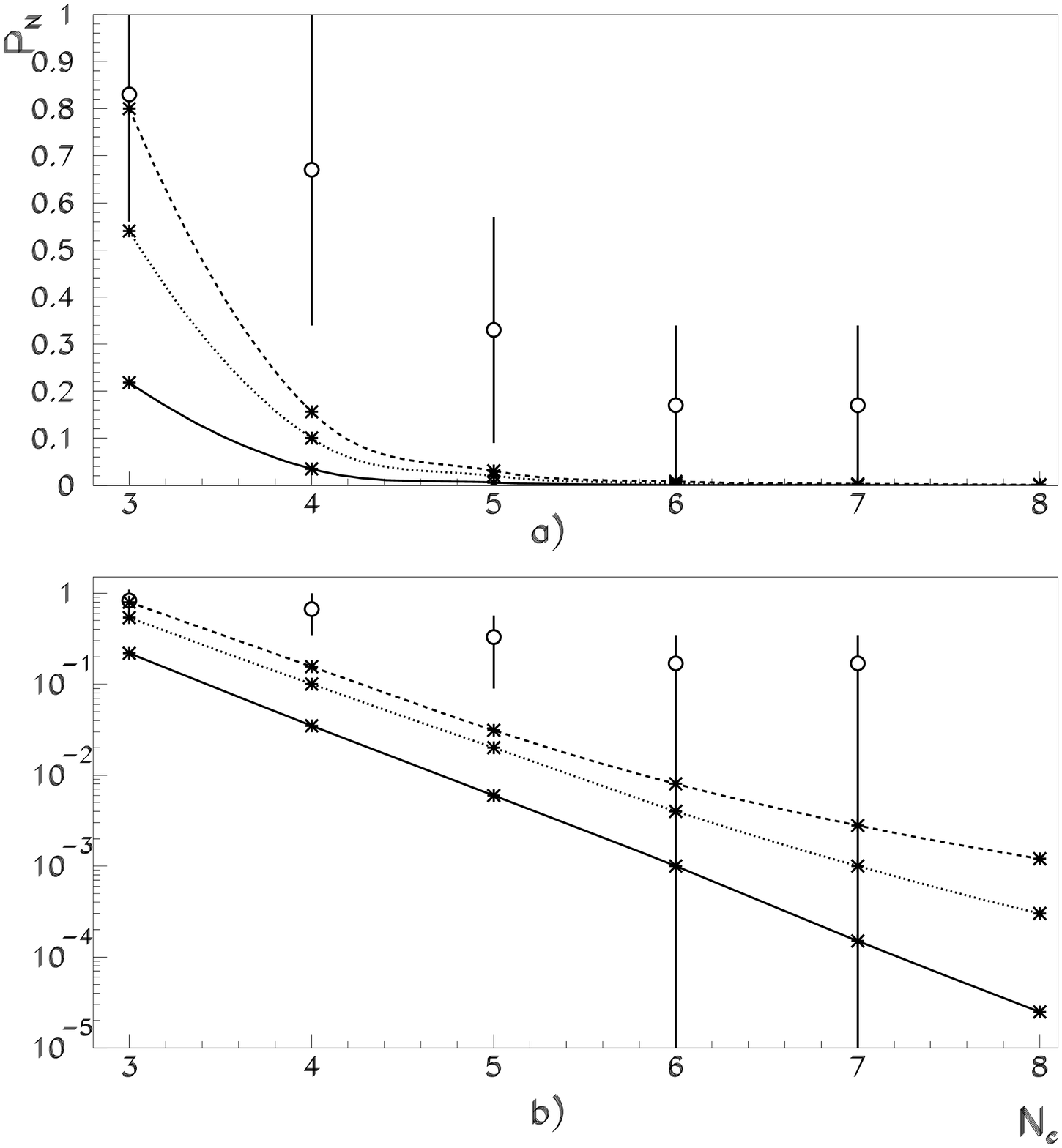, height=220mm}}    
\caption{\small The alignment degree $P_{N_c}$ as a function of cluster number $N_c$ 
at $h=50$ m and  $\sqrt{s}=14$ TeV in linear (a) and logarithmic (b) scales. The 
solid curve is the result (coincident with one at $h=1000$ m) 
without restriction on the total cluster energy
$E^{\rm thr}_{\Sigma}$, the dotted curve --- at $E^{\rm thr}_{\Sigma}=2$ PeV,
the dashed curve --- at $E^{\rm thr}_{\Sigma}=10$ PeV.  Points
($\circ$) with errors are experimental data from~\cite{man}.}
\end{center}
\end{figure}

\begin{figure} 
\begin{center} 
\makebox{\epsfig{file=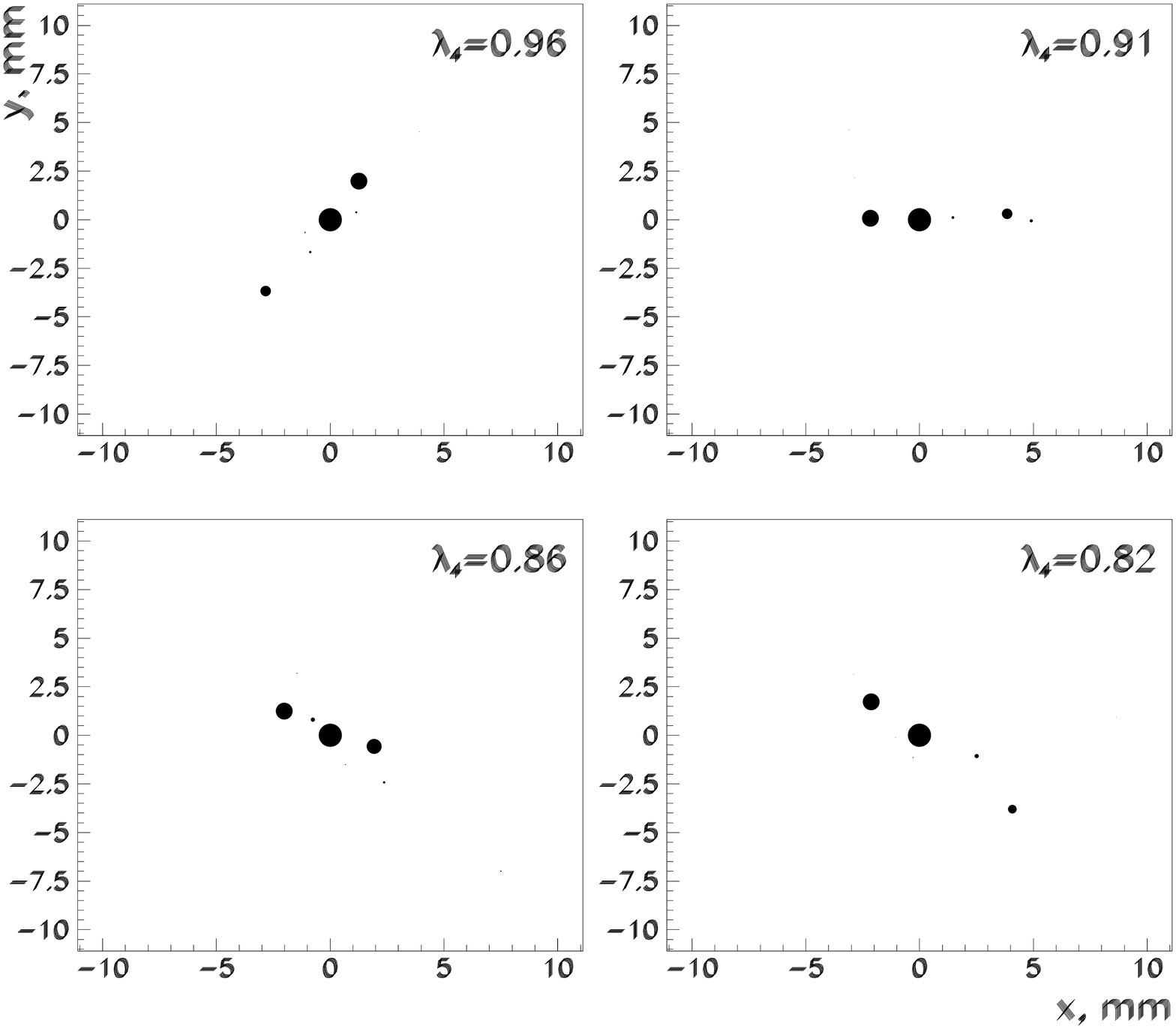, height=180mm}}    
\caption{\small Samples of core distributions for simulated events with 
$E^{\rm thr}_{\Sigma}=10$ PeV and $\lambda_4 > 0.8$.
The size of spots  is proportional to their energy (except for the central 
spot which is not to scale).}
\end{center}
\end{figure}

\begin{figure} 
\begin{center} 
\makebox{\epsfig{file=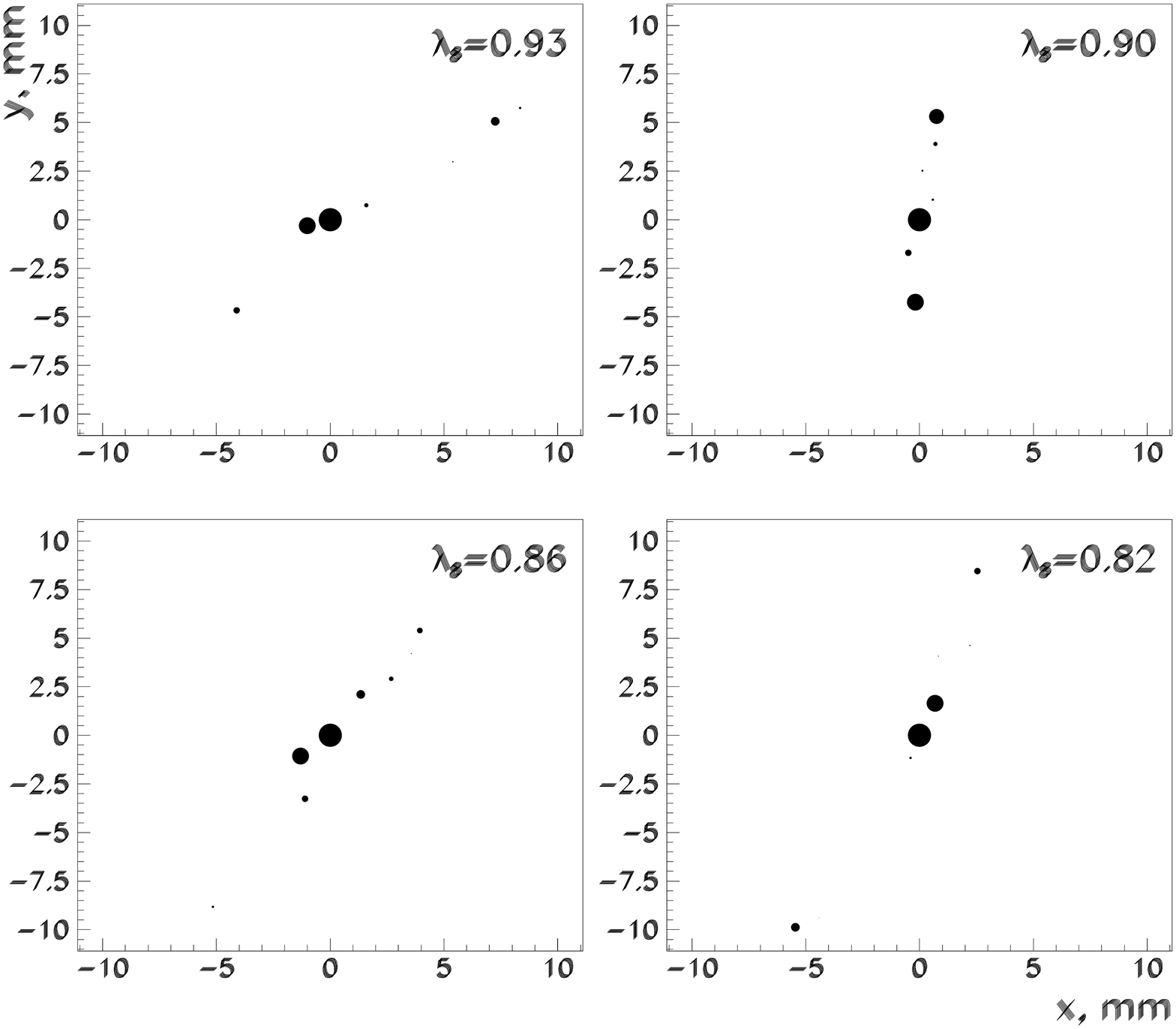, height=180mm}}    
\caption{\small Samples of core distributions for simulated events with 
$E^{\rm thr}_{\Sigma}=10$ PeV and $\lambda_8 > 0.8$.
The size of spots  is proportional to their energy (except for the central 
spot which is not to scale).}
\end{center}
\end{figure}

\begin{figure} 
\begin{center} 
\makebox{\epsfig{file=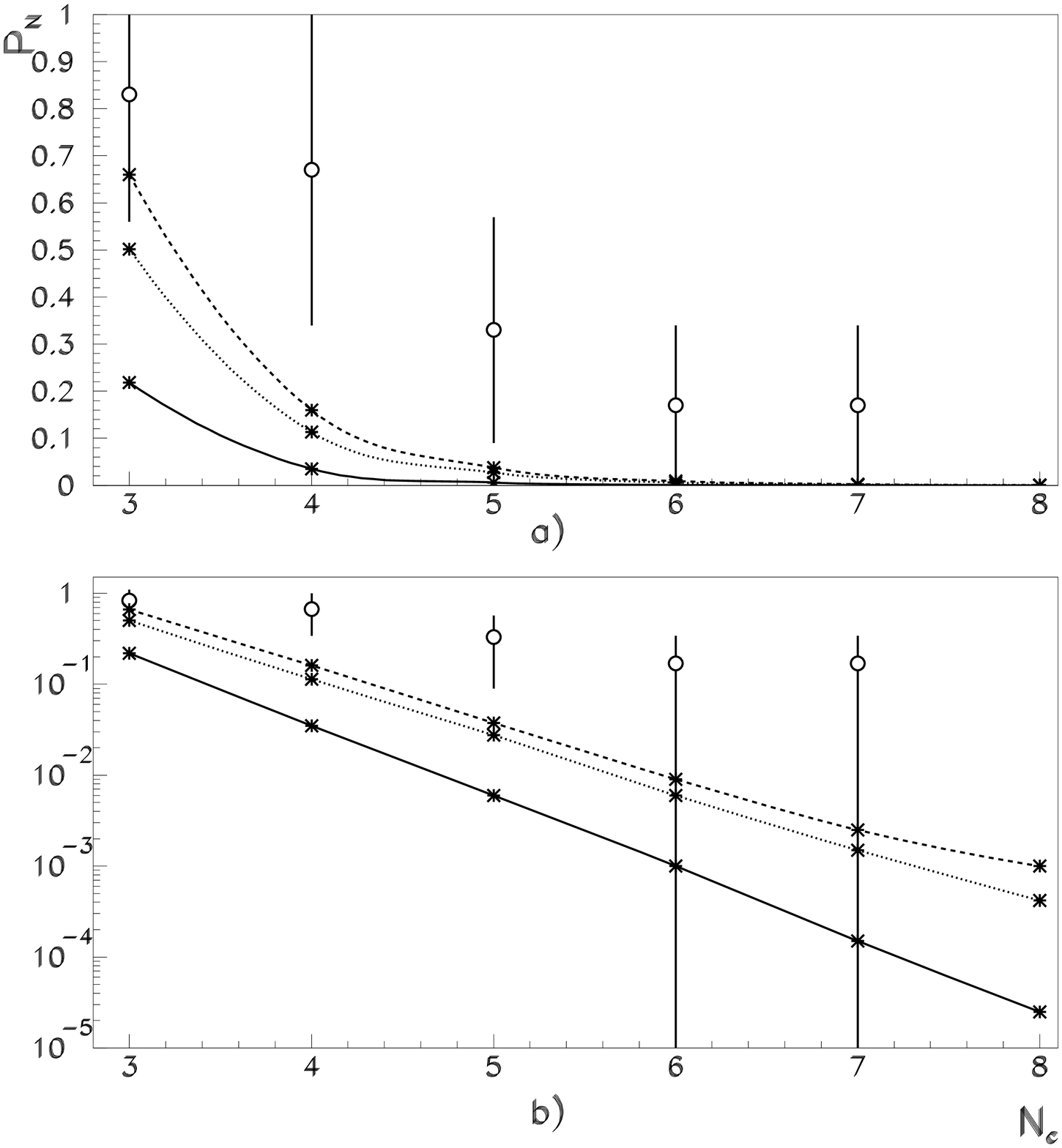, height=220mm}}
\caption{\small The alignment degree $P_{N_c}$ as a function of cluster number $N_c$ 
at $h=50$ m and  $\sqrt{s}=14$ TeV in linear (a) and logarithmic (b) scales. The 
solid curve is the result (coincident with one at $h=1000$ m) 
without restriction on the total  energy of $\gamma$-quanta
$E^{\rm thr}_{\gamma}$, the dotted curve --- at $E^{\rm thr}_{\gamma}=1$ PeV,
the dashed curve --- at $E^{\rm thr}_{\gamma}=5$ PeV.  Points
($\circ$) with errors are experimental data from~\cite{man}.}    
\end{center}
\end{figure}

\end{document}